\documentclass{emulateapj}
\usepackage{multirow}
\usepackage{longtable}
\usepackage{ulem}

\newcommand{\eqref}[1]{Equation (\ref{#1})}
\newcommand{\tabref}[1]{Table~\ref{#1}}
\newcommand{\figref}[1]{Figure~\ref{#1}}

\newcommand{\C}[1]{{}^{#1}{\rm C}}
\newcommand{\0}[1]{{}^{#1}{\rm O}}
\newcommand{\Ne}[1]{{}^{#1}{\rm Ne}}
\newcommand{\Ni}[1]{{}^{#1}{\rm Ni}}
\newcommand{\cdens}{\rho_{c}}
\newcommand{\tcool}{t_{\rm cool}}
\newcommand{\dmb}{\Delta m_{15}(B)}
\shorttitle{SNIa dependence on central density}

\begin{document}

\submitted{Accepted to the Astrophysical Journal Letters}

\title{On variations of the brightness of type Ia supernovae \\
with the age of the host stellar population}

\author{
Brendan K.\ Krueger\altaffilmark{1},
Aaron P.\ Jackson\altaffilmark{1},
Dean M.\ Townsley\altaffilmark{2},
Alan C.\ Calder\altaffilmark{1,3},
Edward F.\ Brown\altaffilmark{4},
F.\ X.\ Timmes\altaffilmark{5}
}

\altaffiltext{1}{
Department of Physics \& Astronomy,
The State University of New York - Stony Brook, Stony Brook, NY
}
\altaffiltext{2}{
Department of Physics and Astronomy,
The University of Alabama, Tuscaloosa, AL
}
\altaffiltext{3}{
New York Center for Computational Sciences,
The State University of New York - Stony Brook, Stony Brook, NY
}
\altaffiltext{4}{
Department of Physics and Astronomy,
Michigan State University, East Lansing, MI
}
\altaffiltext{5}{
School of Earth and Space Exploration,
Arizona State University, Tempe, AZ, USA
}

\begin{abstract}
Recent observational studies of type Ia supernovae (SNeIa) suggest correlations
between the peak brightness of an event and the age of the progenitor stellar
population.  This trend likely follows from properties of the progenitor white
dwarf (WD), such as central density, that follow from properties of the host
stellar population.  We present a statistically well-controlled, systematic
study utilizing a suite of multi-dimensional SNeIa simulations investigating
the influence of central density of the progenitor WD on the production of
Fe-group material, particularly radioactive $\Ni{56}$, which powers the light
curve.  We find that on average, as the progenitor's central density increases,
production of Fe-group material does not change but production of $\Ni{56}$
decreases.  We attribute this result to a higher rate of neutronization at
higher density.  The central density of the progenitor is determined by the
mass of the WD and the cooling time prior to the onset of mass transfer from
the companion, as well as the subsequent accretion heating and neutrino losses.
The dependence of this density on cooling time, combined with the result of our
central density study, offers an explanation for the observed age-luminosity
correlation: a longer cooling time raises the central density at ignition
thereby producing less $\Ni{56}$ and thus a dimmer event.  While our ensemble
of results demonstrates a significant trend, we find considerable variation
between realizations, indicating the necessity for averaging over an ensemble
of simulations to demonstrate a statistically significant result.
\end{abstract}


\section{Introduction}
\label{sec:intro}

Observations targeting the environment of type Ia supernovae (SNeIa) have
exposed open questions concerning the dependence of both their rates and
average brightness on environment.  \citet{MannucciEtAl06} show that the
dependence of the SNIa rate on delay time (elapsed time between star formation
and the supernova event) is best fit by a bimodal delay time distribution (DTD)
with a prompt component less than 1~Gyr after star formation and a tardy
component several Gyr later.  Although the clarity of this effect is clouded by
galaxy sampling~\citep{FilippenkoUnimodalDTD}, the basic result is borne out
even within galaxies~\citep{RaskinEtAl09}.  \citet{GallagherEtAl08} measure a
correlation between the brightness of a SNIa and its delay time, which they
state is consistent with either a bimodal or a continuous DTD.  Other recent
studies by \citet{howelletal+09}, \citet{NeillEtAl09}, and \citet{BrandtEtAl10}
also find a correlation between the delay time and brightness of a SNIa. 

\citet{PhillipsRelation} identified a linear relationship between the maximum
B-band magnitude of a light curve and its rate of decline.  This ``brighter
equals broader'' relationship has been extended to additional bands with
templates from nearby events, allowing SNeIa to be calibrated as an extension
of the astronomical distance ladder~\citep[see][for a description of one
method]{jha2007}.  The brightness of a SNIa is determined principally by the
radioactive decay of $\Ni{56}$ synthesized during the
explosion~\citep{truranetal67,colgatemckee69,arnett:type,
pinto.eastman:physics}.

A widely-accepted proposal to explain many, if not most, events is the
thermonuclear disruption of a white dwarf (WD) in a mass-transferring binary
system~\citep[for reviews from various perspectives see:][]{branchetal1995,
Fili97,hillebrandt.niemeyer:type,livio2000,roepke2006}.  In this paradigm, a
longer delay time suggests the possibility of a longer elapsed time between the
formation of the WD and the onset of accretion.  During this period, denoted
here as the WD cooling time ($\tcool$), the WD is isolated from any 
significant heat input and decreases in temperature.  A longer $\tcool$
results in a higher central density when the core reaches the ignition
temperature~\citep{LesaffreEtAl06}, due to the lower entropy at the onset of
accretion.  Thus, a correlation between central density and the peak brightness
of an event suggests a correlation between delay time and the brightness of an
event.  While previous work indicated a correlation between central density and
peak brightness, none has averaged over a statistically significant ensemble of
realizations~\citep{Bracetal00,Roepetal06_2,hoeetal2010}.  Therefore, we
investigate, for the first time, a statistically significant correlation
between progenitor central density and average peak brightness of SNeIa.

The surrounding stellar population, the metallicity and mass of the progenitor,
the thermodynamic state of the progenitor, the cooling and accretion history of
the progenitor, and other parameters are known to affect the light curves of
SNeIa; the role, and even primacy, of these various parameters is the subject
of ongoing study~\citep[e.g.,][]{Roepetal06_2,hoeetal2010,jacketal10}.
Additionally, many of these effects may be interconnected in complex
ways~\citep{LesaffreEtAl06}.  In this study, we isolate the direct effect of
varying the progenitor central density on the production of $\Ni{56}$.  To
first order, this yield controls the brightness of an event; second-order
effects on the light curve are left for future study.

\section{Method}
\label{sec:method}

Once a WD forms in a binary, it is initially isolated and slowly cools in a
single-degenerate scenario.  Eventually, mass-transfer begins to carry light
elements from the envelope of the companion to the surface of the WD.  If the
accretion rate exceeds a threshold, the infalling material experiences steady
burning~\citep{nomoto_2007_aa}, and eventually the WD gains enough mass to
compress and heat the core.  Once the temperature rises enough to initiate
carbon reactions, the core begins to convect (or ``simmer'').  Our progenitor
models parameterize the WD at the end of this simmering phase, just prior to
the birth of the flame that eventually will disrupt the entire WD in a SNIa.

We constructed a series of five parameterized, hydrostatic progenitor models
that account for simmering in which we vary the central density ($\cdens$).
The outer regions are isothermal, although some temperature structure is
expected~\citep{Kuhletal06}, and the cores are isentropic due to convection and
have a lower C/O ratio~\citep{Straetal03,PiroBild07,Chametal08,PiroChan08}.
\citep{jacketal10} explored these effects and we chose the core composition as
40\% $\C{12}$, 57\% $\0{16}$, and 3\% $\Ne{22}$ and the outer layer as 50\%
$\C{12}$, 48\% $\0{16}$, and 2\% $\Ne{22}$.  For our $\cdens$ we chose
$1-5 \times 10^9$~g/cm$^3$ in steps of $10^9$~g/cm$^3$.  The central
temperature must be in the range of carbon ignition, which is approximately
$7 - 8 \times 10^8$~K~\citep[e.g.,][]{Kuhletal06}; we selected
$7 \times 10^8$~K.  Based on prior research, we chose other model parameters to
produce expected amounts of Fe-group elements in the explosion
\citep{townetal09,jacketal10}.  The values were kept constant in all
simulations in order to isolate the central density effects.

With these five progenitor models, we utilize the statistical framework
presented in \citet{townetal09} for a controlled study of the effect of varying
the central density.  For each progenitor, we created thirty realizations
seeded by a random number used to generate a unique set of spherical harmonics
with power in modes $12 \leq \ell \leq 16$.  The spectra are used as initial
perturbations to the spherical flame surface around the center of the
progenitor star.  Each progenitor uses the same seed values, resulting in the
same thirty perturbations.  This choice allows us to check for systematic
biases in the realizations across different progenitors.

We performed a suite of 150 two-dimensional, axisymmetric simulations of the
deflagration-to-detonation transition (DDT) model of SNeIa with a customized
version of FLASH, a compressible, Eulerian, adaptive-mesh, hydrodynamics code.
The modifications to this code are (1)~the burning model, (2)~the flame speed
computations, (3)~the mesh refinement criteria, (4)~the DDT criterion.  Our
simulation methods are described in detail in previous
publications~\citep{Caldetal07,townsley.calder.ea:flame,townetal09} and
continue to be improved~\citep{jacketal10,townetal10}.  We should note that the
DDT criterion is based on a characteristic density at which we ignite
detonations, which we select to be $10^{7.1}$~g/cm$^3$.  Following the
procedures described in \citet{Caldetal07} and \citet{SeitTownetal09} for
calculating the neutronization rate in material in nuclear statistical
equilibrium (NSE), we utilize weak rates from \citet{FullerEtAl85},
\citet{OdaEtAl94}, and \citet{langanke.martinez-pinedo:weak}, with newer rates
superseding earlier ones.  The reaction networks for calculating the energetics
and time scales of the deflagration and detonation phases included the same 200
nuclides, and the NSE calculation of tables included 443 nuclides. The
supernova simulations use a three-stage burning model in which the timescale to
burn to NSE is calibrated to reproduce the correct yield of Fe-group elements.
The weak reaction rate is negligible at a density where Si-burning to $\Ni{56}$
is incomplete; therefore, we estimate the $\Ni{56}$ yield from the total NSE
yield and its associated electron fraction.  We performed nuclear
postprocessing of Lagrangian tracer particles using a network of 200 nuclides
on a subset of simulations.  Comparison demonstrates that our burning model
matches the energetics and final composition of important nuclei such as
$\Ni{56}$~\citep{townetal10} \citep[see also studies by][]{travetal+04,
sietetal+10}.  We utilize the adaptive-mesh capability, with a highest
resolution of 4~km, which demonstrates a converged result~\citep{townetal09}.

\section{Results and Discussion}
\label{sec:results}

The $\Ni{56}$ yield is determined partly by neutronization occurring during the
thermonuclear burning.  Neutronization pushes the nucleosynthetic yield away
from balanced nuclei such as $\Ni{56}$ to more neutron-rich, stable isotopes
like $\Ni{58}$.  Thus the amount of neutronization influences the brightness of
an event and, all else being constant, more neutronization results in a dimmer
event.  The degree of neutronization depends on the density and temperature
evolution of burned material.  Generally, thermonuclear burning occurring at
higher densities will neutronize faster.  In an explosive event like a
supernova, the longer material remains in NSE at high densities, the more
neutronization occurs~\citep{Nomo84,khokhlov91b,Caldetal07}.  Accordingly, for
SNeIa, both the central density and the duration of the deflagration phase
influence the brightness of an event.

\begin{figure}[t]
\includegraphics[angle=270,width=\columnwidth]{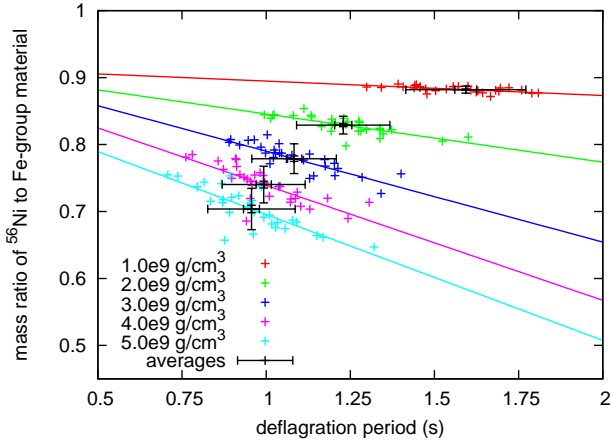}
\caption{\label{fig:time}Plot of the $\Ni{56}$-to-NSE mass ratio vs.\ duration
of deflagration.  Small crosses are single simulations and are colored by
$\cdens$.  Large black crosses show the average values for a given $\cdens$
with error bars showing the standard deviation and the standard error of the
mean.}
\end{figure}

\figref{fig:time} presents the mass fraction of NSE material that is $\Ni{56}$
as a function of deflagration duration, with points colored to indicate
$\cdens$.  The duration of the deflagration phase is the time elapsed between
the formation of a flame front and the ignition of the first detonation point.
We consider this elapsed time because there is little contribution to
neutronization after the first DDT.  The mass of NSE material produced
increases during the course of the SNIa and eventually plateaus; we find the
point where the NSE yield changes by less than 0.01\% over 0.01~s and use that
mass as the final yield of the SNIa.  The results have considerable scatter but
show two trends.  First, at a given $\cdens$, simulations with longer
deflagration periods tend to have a greater degree of neutronization.  Next,
simulations from progenitors with higher central densities tend to have a
greater degree of neutronization despite having shorter deflagration periods.
This result shows that the increased rate of neutronization at higher
densities, which can be seen by the steeper slopes of the higher density trend
lines, more than compensates for the decrease in time for neutronization to
occur.  Accordingly, our results qualitatively agree with previous
work~\citep[e.g.,][]{Iwametal99,Bracetal00,hoeetal2010}.  We also note that the
yield of NSE material is, within the error of the slope, independent of
$\cdens$.  As seen in \citet{WoosleyEtAl07:LightCurves}, if the NSE yield
remains constant but the amount of $\Ni{56}$ varies, then the results should
lie along the Phillips relation.

\begin{figure}[t]
\includegraphics[angle=270,width=\columnwidth]{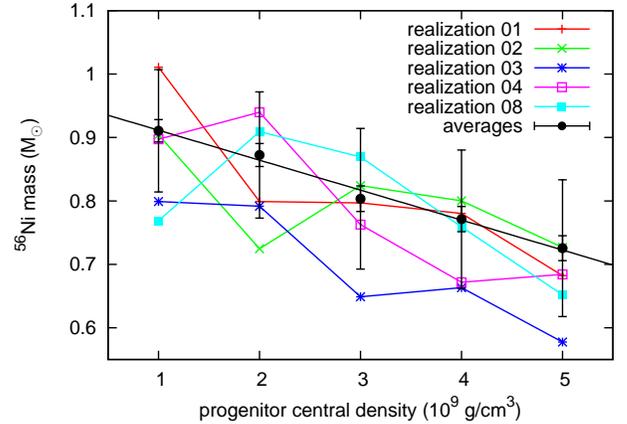}
\caption{\label{fig:dens}Plot of mass of $\Ni{56}$ produced vs.\ $\cdens$ for
five different realizations (colored curves), demonstrating the variety of
trends seen for a single realization.  These include non-monotonic trends,
which could suggest an increase of $\Ni{56}$ with increasing $\cdens$ instead
of the decrease seen in the ensemble.  In black are the average values for each
density, along with the standard deviation, the standard error of the mean, and
a regression fit to the average values.}
\end{figure}

The greater degree of neutronization seen in \figref{fig:time} leads directly
to lower $\Ni{56}$ yields with increasing $\cdens$.  This result can be seen in
\figref{fig:dens}, which presents the $\Ni{56}$ yield of each simulation
plotted against $\cdens$ with color used to classify the simulations by
realization.  The scatter among different realizations at the same $\cdens$ is
greater than the variation across the $\cdens$ range, indicating the need for
analysis of an ensemble of realizations.  This scatter can result in a single
realization showing a trend unlike the statistical trend; for example, by
considering only realization 2 and $\cdens$ of $2 \times 10^9$ and
$3 \times 10^9$~g/cm$^3$ (\figref{fig:dens}, green curve), we would conclude
that increasing $\cdens$ causes an increase in $\Ni{56}$ production, instead of
a decrease as seen in the overall ensemble.  The scatter follows from a strong
dependence on the morphology of the flame surface during the early
deflagration, which varies the duration of the deflagration and the production
of $\Ni{56}$.  Changing $\cdens$ can cause a local change in the plume dynamics
that overrides the general trend.  By fitting the averages we find the relation
\begin{equation}
M_{\Ni{56}} = A \cdens + B,
\end{equation}
where
\begin{eqnarray*}
A & = & -0.047 \pm 0.003 \:\: \frac{M_\odot}{10^9 \rm{g/cm}^3} \\
B & = &  0.959 \pm 0.009 \:\: M_\odot.
\end{eqnarray*}

We use data from \citet{LesaffreEtAl06} to correlate $\cdens$ to $\tcool$, one
component of the delay time, specifically the results for a WD with a
pre-accretion mass of $1$~$M_\odot$.  Thus if we imagine a collection of stars
forming with the same zero-age main-sequence mass but with different companions
and binary separations, the delay time would be dominated by $\tcool$.  There
are large uncertainties in the relationship between $\cdens$ and $\tcool$;
accordingly, we neglect error derived from this large uncertainty in our
analysis.  We note that the work of \citet{LesaffreEtAl06} suggests that a WD
with a central density of $10^9$~g/cm$^3$ will not ignite; further accretion is
necessary to reach ignition conditions.  Therefore we cannot use
\citet{LesaffreEtAl06} to compute a $\tcool$ for our lowest-$\cdens$
simulations, and so we omit such simulations.

\begin{figure}[t]
\includegraphics[width=\columnwidth]{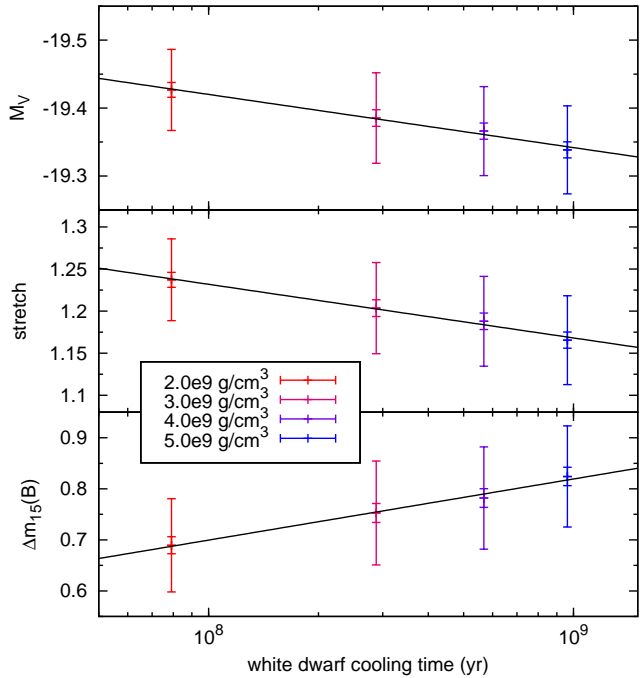}
\caption{\label{fig:obs}Plot of brightness vs.\ $\tcool$ in terms of $M_V$
(upper panel), $s$ (central panel), and $\dmb$ (lower panel).  Plotted are the
average values for each $\cdens$ along with the standard deviation, the
standard error of the mean, and a regression fit to the average values.}
\end{figure}

We convert our $\Ni{56}$ mass to stretch ($s$)~\citep{howelletal+09}, the
magnitude difference in the B band between maximum light and 15 days after
maximum light ($\dmb$)~\citep{goldhaberetal01}, and absolute magnitude in the V
band at maximum light ($M_V$)~\citep{phillips.lira.ea:reddening-free} in order
to compare with observational findings, shown in \figref{fig:obs}.  A detailed
comparison of our models to these observables requires radiative transport
calculations of synthetic spectra and light curves, but our results drawn from
the $\Ni{56}$ mass are sufficient for the basic properties we consider here.
\citet{WoosleyEtAl07:LightCurves} showed that the brightness of an event
depends on the mass of $\Ni{56}$ when it is distributed through a large
fraction of the star as in our simulations.  Additionally, \citet{StritzEtAl06}
showed that late-time nebular spectroscopy finds $\Ni{56}$ yields consistent
with those found from peak luminosity using the inverse of the relations we
use.

We find that the best-fit relations follow the form
\begin{equation}
q = \alpha_q \, \log_{10} \left( \frac{t_{\rm cool}}{yr} \right) + \beta_q
\label{eqn:trend}
\end{equation}
where $q$ is one of $\dmb$, $s$, or $M_V$.  The values for $\alpha_q$ and
$\beta_q$ are shown in \tabref{tab:fit}.

\begin{table}
\centering
\caption{\label{tab:fit}Best-fit parameters for brightness-age relations.}
\begin{tabular}{c c c}
\hline \hline
$q$ & $\alpha_q$ & $\beta_q$ \\
\hline
$M_V$  & $0.078  \pm 0.006$ & $-20.05 \pm 0.05$ \\
$s$    & $-0.064 \pm 0.005$ & $1.74   \pm 0.04$ \\
$\dmb$ & $0.120  \pm 0.009$ & $-0.26  \pm 0.08$ \\
\hline \\
\end{tabular}
\end{table}

\begin{figure}[t]
\includegraphics[width=\columnwidth]{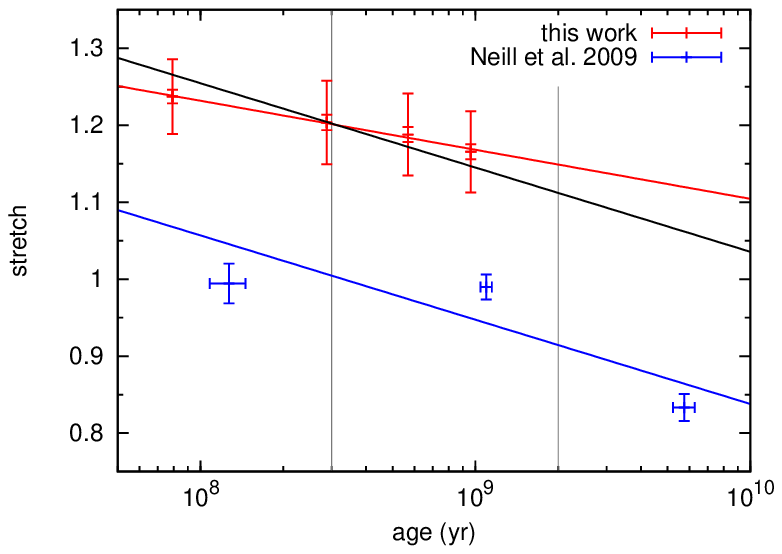}
\caption{\label{fig:neill}Plot of stretch vs.\ age.  In red are the points from
this study, based on variations in $\cdens$, along with the standard deviation,
the standard error of the mean, and a best-fit trend line following the form of
\eqref{eqn:trend}.  In blue are the binned and averaged points from Figure 5 of
\citet{NeillEtAl09}, along with a best-fit trend line following the form of
\eqref{eqn:trend}.  The vertical grey lines mark the cuts between bins in the
\citet{NeillEtAl09} analysis.  The trend line for the \citet{NeillEtAl09} data
is shifted upward for comparison to our results (black line).  The overall
offset to larger stretch in the simulations is due to the choice of DDT
density.  The approximate agreement of the overall trend indicates that the
variation of $\cdens$ is an important contributor to the observed trend, but
that other factors are also important.}
\end{figure}

To highlight the comparison with observations, in \figref{fig:neill} we plot an
expansion of the center panel from \figref{fig:obs} with the binned results
from Figure 5 of \citet{NeillEtAl09}.  While an absolute comparison is not
possible, the similarity of the overall trend indicates that variation of
$\cdens$ is an important contribution to the observed dependence.  Our choice
of initial conditions and DDT density results in an effective calibration that
yields higher than expected $\Ni{56}$ masses.  Accordingly, our results are
systematically too bright, giving abnormally high values of stretch.  Future
studies will correct for this effect.  A more subtle point comes in the usage
of ``age'': \citet{NeillEtAl09} measures the luminosity-weighted stellar age,
while for the theory we have simply used $\tcool$ directly, which for late
times is the dominant portion of the time elapsed since star formation.  Such
offsets, either vertical or horizontal, are less important than the overall
trend and the range that can be attributed to variation of $\cdens$.  For
comparison, the black line in \figref{fig:neill} shifts the best-fit line from
the data of \citet{NeillEtAl09} up to align it with our results.  The trend due
to $\cdens$ is weaker than in observations, suggesting that $\cdens$
contributes to the observed trend but that other effects also play a part.

\section{Conclusions and Future Work}
\label{sec:conclusions}

We simulated a suite of 150 SNIa models with a range of $\cdens$ to study the
trends for a population of SNeIa.  We find that on average progenitors with
higher $\cdens$ produce less $\Ni{56}$.  \citet{hoeetal2010} argue that
$\Ni{56}$ in the central regions of the exploding WD does not contribute to the
light curve at maximum, and therefore they do not see a significant trend with
central density in the maximum V-band magnitude, but rather in late-time
brightness.  The pure-deflagration models of \citet{Roepetal06_2} exhibit a
shallow increase of produced $\Ni{56}$ as central density increases, in
contradiction of our findings.  \citet{Iwametal99} find that the trend with
central density depends on the DDT transition density; extrapolating from their
results, our value of $\rho_{\rm DDT}$ should yield an increasing $\Ni{56}$
yield as central density increases.  We find that small perturbations of the
initial flame surface not only influence the final $\Ni{56}$ yield, but also
its dependence on central density through variations in the duration of the
deflagration phase caused by differences in plume development.  The variation
that follows from perturbations on the initial conditions is a critical aspect
of multi-dimensional modeling.  Only after many realizations with different
perturbations of the initial flame surface are simulated does a statistically
significant trend with central density emerge.  This result, illustrated by
\figref{fig:dens}, demonstrates the need for an ensemble of simulations to
explore systematic effects in SNeIa.

By relating $\cdens$ to $\tcool$ and $\Ni{56}$ to $\dmb$, our results support
the observational finding that SNeIa from older stellar populations are
systematically dimmer.  While a degeneracy between age and metallicity in the
integrated light of stellar populations exists, the observed dependence of mean
brightness of SNeIa on mean stellar age is apparently the stronger
effect~\citep{GallagherEtAl08,howelletal+09}.  Accordingly, our choice to
neglect metallicity effects and consider only the effect of central density on
$\Ni{56}$ yield allows us to offer a theoretical explanation for this observed
trend.  If we additionally consider the effect of metallicity, we may see a
slightly stronger trend of decreasing brightness with increasing age as has
been previously suggested~\citep{timmes.brown.ea:variations}.  Other effects
besides progenitor central density and metallicity, such as progenitor main
sequence mass, may also contribute to this trend.

The insensitivity of the overall Fe-group yield to central density, and
therefore delay time, along with the dependence of the $\Ni{56}$ yield on
central density, implies that SNeIa of similar brightnesses (and therefore
similar $\Ni{56}$ yield) from progenitors of different ages will not have the
same total Fe-group yield.  Those from older populations will, on average, have
larger masses of stable species.  This may argue for a slight non-uniformity in
the Phillips relation based on environment~\citep{WoosleyEtAl07:LightCurves,
hoeetal2010}.  The resulting closely-related family of brightness-decline time
relations also provides a physical motivation for intrinsic scatter in the
Phillips relation as a result of combining populations with different mean
stellar ages.  In this picture the primary parameter is the degree of expansion
at DDT, determined by the morphology of the early flame (and the DDT density,
which we hold constant), and the age acts a weaker secondary parameter.  In any
case, the possibility of such an effect motivates further exploration of the
impact of central density on the light curve itself.

\acknowledgements

This work was supported by the Department of Energy through grants
DE-FG02-07ER41516, DE-FG02-08ER41570, and DE-FG02-08ER41565, and by NASA
through grant NNX09AD19G.  ACC also acknowledges support from the Department of
Energy under grant DE-FG02-87ER40317.  DMT received support from the Bart
J.\ Bok fellowship at the University of Arizona for part of this work.  The
authors gratefully acknowledge the generous assistance of Pierre Lesaffre, as
well as fruitful discussions with Mike Zingale, and the use of NSE and weak
reaction tables developed by Ivo Seitenzahl.  The authors also acknowledge the
hospitality of the KITP, which is supported by NSF grant PHY05-51164, during
the programs ``Accretion and Explosion: the Astrophysics of Degenerate Stars''
and ``Stellar Death and Supernovae.''  The software used in this work was in
part developed by the DOE-supported ASC/Alliances Center for Astrophysical
Thermonuclear Flashes at the University of Chicago.  This research utilized
resources at the New York Center for Computational Sciences at Stony Brook
University/Brookhaven National Laboratory which is supported by the
U.S.\ Department of Energy under Contract No.\ DE-AC02-98CH10886 and by the
State of New York.

\end{document}